\documentclass[]{acm_sen_article}

\usepackage{lipsum}
\usepackage{flushend}

\usepackage{graphicx} 
\PassOptionsToPackage{hyphens}{url}
\usepackage{hyperref}
\usepackage{xspace}
\usepackage[numbers]{natbib}
\usepackage{balance}
\usepackage{xcolor}
\hypersetup{
  colorlinks=false,
  pdfborder={0 0 0}
}

\newcommand{\ie}{\emph{i.e.,}\xspace}
\newcommand{\eg}{\emph{e.g.,}\xspace}
\newcommand{\etc}{\emph{etc.}\xspace}

\usepackage{fontawesome5}
\usepackage{framed}
\definecolor{formalshadelight}{RGB}{242,242,242}
\definecolor{formalshadedark}{RGB}{166,166,166}
\newenvironment{highlight}{%
  \MakeFramed{\advance\hsize-\width\FrameRestore}%
  \noindent\begin{minipage}{\linewidth}\noindent\hspace{-4.55pt}
  \vspace{2pt}
}
{%
  \vspace{2pt}\vspace{-2pt}\end{minipage}\endMakeFramed%
}

\usepackage{csquotes}

\begin{document}

\title{Notes On Writing Effective Empirical Software Engineering Papers: An Opinionated Primer}

\numberofauthors{2} 
\author{
Roberto Verdecchia\\
       \affaddr{University of Florence}\\
       \affaddr{Italy}\\
       \email{roberto.verdecchia@unifi.it}
\and
Justus Bogner\\
       \affaddr{Vrije Universiteit Amsterdam}\\
       \affaddr{The Netherlands}\\
       \email{j.bogner@vu.nl}
}

\maketitle


\section{Introduction}
While mastered by some, good scientific writing practices within Empirical Software Engineering (ESE) research appear to be seldom discussed and documented.
Despite this, these practices are implicit or even explicit evaluation criteria of typical software engineering conferences and journals.
In this pragmatic, educational-first document, we want to provide guidance to those who may feel overwhelmed or confused by writing ESE~papers, but also those more experienced who still might find an opinionated collection of writing advice useful.
The primary audience we had in mind for this paper were our own BSc, MSc, and PhD students, but also students of others.

If you have made it this far, you probably do not need much convincing why good scientific writing is important, but let us briefly cover it anyway.
Most people will (hopefully) agree that it is even more important to conduct rigorous research.
However, if you cannot communicate your research properly, it is at best not living up to its
potential and, at worst, useless.
Additionally, good scientific writing makes it easier for reviewers to identify solid research, and it may convince them that decent research is \enquote{good enough} to be accepted.
Bad scientific writing, on the other hand, may demotivate or even anger your reviewers, making acceptance less likely.
It may give the impression of incompetence, sloppiness, or laziness, all of which you definitely want to avoid.
Ideally, reviewers should focus much more on the research than on its presentation.
However, many ESE venues have an explicit review criterion related to good scientific writing (\enquote{presentation}), and it plays an even larger role subconsciously.
We therefore think that every researcher should at least try to improve their writing abilities, and that we as the ESE community should strive for papers that are easy to understand and enjoyable to read.

As a warning note to the reader, this paper follows what could be deemed as a rather unconventional structure.
The subsection names of Section~\ref{sec:sections} are template names of common ESE sections, and not the sections of the paper itself.
To ease distinguishing them from the actual sections of this paper, all exemplary template names are preceded by the prefix \enquote{The}, \eg \enquote{The Introduction} instead of the actual \enquote{Introduction} you are reading now.

\textbf{Disclaimer:} While we have both attended several courses on scientific writing, the following guidance and tips are also partially derived from personal experience of working on and supervising ESE paper writing and from discussions with colleagues.
Our documented advice therefore reflects a subjective and personal vision of writing ESE papers.
By no means do we claim to be fully objective, generalizable, or representative of the whole discipline.
With that being said, writing papers in this way has worked pretty well for us so far.
We hope that this guide can at least partially do the same for others.

\section{A Typical ESE Paper Structure}
While ESE papers can vary in their structure, most usually contain -- with potentially different names and order -- the following sections.
For more information on the content of each section, please refer to Section~\ref{sec:sections}.

\begin{itemize}
    \item \textbf{Abstract:} a concise and precise summary of the whole paper that covers the key points of all major sections (think of it as an executive summary for researchers);
    \item \textbf{Introduction:} the general context of the study, the motivation for the study, \ie the importance of the problem to be solved, an outline of the solution / observations, the main contributions of the study (sometimes combined with the main results), and potentially the intended readers;
    \item \textbf{Background:} provides not widely known information required to understand the study (if necessary), \eg important foundational concepts;
    \item \textbf{Related Work:} a discussion of the related work, used to position the study in its academic and / or industrial context (what is new / different?), ideally with a summary at the end;
    \item \textbf{Approach:} if the paper is primarily about a design contribution, then this sections provides a description of the proposed approach (if not, you likely do not need it);
    \item \textbf{Study Design:} declaration of the used research method(s) and detailed description of the study design, execution, and analysis, starting from the research goal and research questions (RQs), to a step-by-step description of the research process, without talking about the outcomes (sometimes also called \enquote{Research Method}, \enquote{Research Design}, \etc);
    \item \textbf{Results:} fairly objective and neutral presentation of the results, typically grouped by the RQs (subsections);
    \item \textbf{Discussion:} discussion of the results, including explanations and reasonable conjectures about their meaning (interpretation), their implications (potentially for different stakeholders), and optionally other lessons learned or takeaways not strongly related to the RQs;
    \item \textbf{Threats to Validity:} discussion of potential threats to the validity of the study, usually grouped into different threat categories;
    \item \textbf{Conclusion:} very brief summary of the study and the findings, followed by opportunities for future research based on this work;
    \item \textbf{Acknowledgments:} brief acknowledgements of non-author contributions and funding schemes (if any);
\end{itemize}

\section{Content of Common ESE Sections}
\label{sec:sections}
In the following, we describe the content of each section and provide tips on how to write it.
At the end, we specify the typical length of the section, calculated by considering a standard 10+2-page double-column research paper.
Be aware that the structure, the names of the sections, and also their content can change strongly depending on the used research methods and study contributions.
While we try to provide a general summary, there are also excellent, more specific resources for reporting the results of a certain research method, e.g., for controlled experiments~\cite{Jedlitschka2008} or systematic literature reviews~\cite{Budgen2018,kitchenham_segress_2023}.

\subsection{The Title}
The title might be one of the most important parts of the paper, as readers will in most cases decide to read the paper or not based only on the title.
The title should be carefully engineered by evaluating different options and discussing them with co-authors.

A good title needs to balance brevity with being precise and indicative of the paper content.
In terms of length, as a rough estimate, we suggest trying to keep the title under 15 words in total, including short articles, conjunctions, and prepositions.
Regarding its meaningfulness, the title should include the most relevant, commonly used keywords that characterize the paper.

Usually, the title is a combination of keywords that identify (i) the general topic, (ii) the content of the work, and potentially (iii) the core contribution of the work that makes the research stand out.
As an example, let us consider a paper that presents a family of algorithms for test case prioritization.
The algorithms are based on the similarity of test cases, and their core result is a strong improvement in performance.
The title of the paper should include keywords related to \enquote{test case prioritization}, \ie the topic, \enquote{similarity algorithms}, \ie the content of the work, and \enquote{performance}, \ie the core contribution.
Following this process, we created the following title: \enquote{FAST Approaches to Scalable Similarity-based Test Case Prioritization}~\cite{miranda2018fast}.
However, it is definitely possible to create effective shorter titles that only include two of these three concepts, \eg the considered topic and the core result, as is the case for \enquote{When and Why Your Code Starts to Smell Bad}~\cite{tufano2015and}.

Another effective practice to help readers judge the content of a paper is to include the used research method in the title, \eg \enquote{Architectural Technical Debt: A Grounded Theory}~\cite{verdecchia2020architectural} or \enquote{How Do Microservice API Patterns Impact Understandability? A Controlled Experiment}~\cite{bogner_how_2024}.
This will immediately set a clear frame of reference for experienced readers to ground their expectations.
It is particularly effective if the study is mainly based on a single research method.

As a final and more advanced recommendation, titles should ideally also be memorable.
This allows readers to recall the paper more easily and to find it more efficiently when searching for it.
A widespread practice to achieve this goal is to include an adaptation of a common quote, idiom, slogan, or wordplay, \ie something that resonates with the reader or which they might find funny.
Some examples are:

\begin{itemize}
    \item \enquote{How Bugs are Born: A Model to Identify How Bugs are Introduced in Software Components}~\cite{rodriguez2020bugs}
    \item \enquote{To Type or Not to Type? A Systematic Comparison of the Software Quality of JavaScript and TypeScript Applications on GitHub}~\cite{bogner_type_2022}
    \item \enquote{Power Hungry Processing: Watts Driving the Cost of AI Deployment?}~\cite{luccioni_power_2024}
    \item \enquote{High Expectations: An Observational Study of Programming and Cannabis Intoxication}~\cite{he_high_2024}
\end{itemize}

However, be aware that there are also people in the ESE community who do not like \enquote{funny} titles, so best not overdo it.

\subsection{The Abstract}
After the title, the abstract is usually the next part of the paper readers will scan for deciding if a paper is worth reading.
Therefore, abstracts should be of the highest possible quality and clearly summarize the entire paper in a few paragraphs.
Given this summarizing nature, an abstract should be written last when the key points of each section are final (see also Section~\ref{sec:writing_order}).
A good practice while writing them is hence to go from one major section to the next by condensing them in a few sentences.

Abstracts can be \textit{structured}, \ie following a set of predefined, labeled paragraphs.
While the concrete labels can vary slightly, a typical structured abstract of an ESE paper has the paragraphs \enquote{Context}, \enquote{Objectives}, \enquote{Methods}, \enquote{Results}, and \enquote{Conclusions}.
Not all venues require structured abstracts, but we suggest following their ordered content regardless, even if you prefer omitting the labels at the beginning of each paragraph.
Adhering to these paragraphs covers all key parts that a good abstract should include.
All paragraphs should be written in a very precise and concise manner. Avoid delving deeply into any topic, while still providing sufficient information to let the reader understand the content of the entire paper at a glance.

\textit{Context.} The context (sometimes also called \enquote{background} or \enquote{motivation}) lets the reader understand the current state of the general research topic and conveys why the specific paper topic is important to be studied.
In other words, this paragraph briefly summarizes the state of the art / practice and describes the research gap the paper addresses.
For example, in a paper focusing on SE for video games like the one by \citet{zamorano2024game}, the context could explain the special characteristics of this type of SE and the lack of studies focusing on it.

\textit{Objectives.} As a logical extension of the previously defined gap, this paragraph (sometimes also called \enquote{aims)} explicitly states in very few sentences (usually up to three) the goals of the research, \ie the study purpose.
The goal(s) should be fairly high-level to include the whole study, while omitting potential details like a plethora of fine-grained sub-RQs.
If your study only has two or three RQs that can be expressed in short phrases, you can also include them all here, but not in their full original question form (see \citet{bogner_restful_2023} for an example).

\textit{Methods.} The methods paragraph should provide the reader with a condensed summary of the applied research process for data collection and analysis.
This includes (i) the used research method(s), \eg repository mining, controlled experiment, or focus group, (ii) the key design decision per research method, \eg dependent and independent variables, used datasets, or applied participant sampling, and (iii) how the data was analyzed, \eg via manual coding or with certain statistical methods.
Include concrete numbers like sample sizes to allow readers to quickly judge the extensiveness of your study, \eg \enquote{Our investigation evaluates the usability of 290 tests generated by GitHub Copilot for 53 sampled tests from open source project}~\cite{el2024using}.

\textit{Results.} The results paragraph summarizes the main research outputs.
While likely true for every abstract paragraph, this one may require multiple iteration, as it can be difficult to identify and precisely summarize the main study results in a few sentences.
This paragraph should not present \textit{all} study results, but only the ones deemed most important, which can be especially difficult for qualitative or secondary studies.
Try to support key results with concrete numbers as much as possible, \eg \enquote{SparseCoder is four times faster than other methods measured in runtime, achieving a 50\% reduction in floating point operations per second}~\cite{yang2025sparsecoder}.
This allows readers to understand the main study outcomes in a precise and objective way.

\textit{Conclusions.} The last abstract paragraph usually summarizes the key takeaways like study implications, which can also be of a more speculative nature (see also Section~\ref{sec:discussion}).
The conclusions can also hint at what the study outcomes mean for future research and practice, \eg \enquote{Our results suggest that sharing strategic knowledge differs from sharing code and raises challenging questions about how knowledge-sharing platforms should support search and feedback}~\cite{arab2022exploratory}.

\textbf{Length:} Usually 0.25 to 0.5 columns

\subsection{The Introduction}
\label{sec:intro}
The introduction serves to outline the general context of the study, and provides the reader with a swift understanding of the paper content.
In addition to that, its main function is to \textit{motivate} the paper, \eg by explaining why the problem to be solved is important.
A reader should ideally believe in this before they reach your stated objective, so build a clear chain of arguments from the context via the problem to your objective, all supported by appropriate references.
Furthermore, do not only say that nobody has done it before.
There are a great many things that nobody has done before, and it is often better that way.
Why is it \textit{important} to do it, and what do we have to gain from it?
Compare this also with Ipek Ozkaya's advice\footnote{\url{https://x.com/ipekozkaya/status/1734783199002124767}. Accessed 2025-03-20.}: resist the urge to claim \enquote{we are the first to do x}.
This is not only quite hard to validate, but it is also a longitudinal observation that mostly has merit if it comes from others.

A typical Introduction may be composed of the following paragraphs, but especially the later ones are not all mandatory:
\begin{itemize}
    \item Concise introduction to the topic, which explains the need and relevance of the paper. This first part can mention some related work if needed to describe the research context, while leaving a proper discussion of the related work to the dedicated section.
    \item A very brief description of the main contribution of the research, \eg a new approach, theory, conducted empirical study, or replication.
    \item A brief mention of the used research method(s), allowing the reader to gain a high-level overview of how the study was conducted.
    \item A brief summary of the main results of the work, if possible reporting explicitly the results with precise numbers (\eg \enquote{\textit{the approach improves X by Y\%}}).
    \item A bullet list summarizing all major contributions of the work (\eg a novel approach, a conducted evaluation experiment, and a reusable dataset).
    \item The intended readers and what they can do with the results, \eg researchers focusing on a specific area, practitioners wishing to solve a specific problem, SE educators in a specific context, \etc
    \item A clickable link to the replication package (can also be provided elsewhere)
    \item For complex papers, an outline of the paper structure. Otherwise, skip this. You can use the space for something more valuable.
\end{itemize}

\textbf{Length:} Usually less than 2 full columns (1 page).

\subsection{The Background}
\label{sec:background}
If the content of the paper requires non-trivial and / or not widely known information to be understandable, it is recommended to use a dedicated Background section.
However, this section is not required, and its content can also be included in the first paragraphs of the Introduction (see Section~\ref{sec:intro}), provided that this content is short enough.
Commonly, this section presents concepts, terms, tools, formalizations, syntax, \etc on which the study is based.
Therefore, a Background section has the purpose to introduce and explain foundations that are important for the rest of the paper.
Its goal is to improve a reader's understanding, which is also why it should be placed after the introduction.
Some people use the term \enquote{Literature Review} to refer to the Background or even the Related Work section, but we strongly discourage this due to the overloaded nature of the term, e.g., regarding systematic literature studies.

The content of the Background section should follow an orderly structure, \eg by presenting background literature in temporal order, documenting concepts at an increasing level of complexity, and group related concepts in the same / subsequent paragraph(s) or subsections.

Note: The Background section should \textbf{not} discuss the related work.
This difference will be explained more thoroughly below.

\textbf{Length:} Usually less than 2 columns.

\subsection{The Related Work}
The Related Work section in ESE papers is primarily used to defend the novelty and significance of the provided contribution(s).
As such, it is important to carefully discuss the majority of studies that are closely related to your own study.
While the Introduction is the primary place to motivate your study, the Related Work section also plays an important role.
This time, you will present similar work in the area and state how your own study differs from it or extends it, ideally with a short summarizing paragraph at the end of the section that highlights the gap you are filling.
The section can open with a short repetition from the introduction of why the problem is important to address, followed by describing how this problem has been (partially) addressed in the past.

\subsubsection{Related Work vs. Background}
So, how does this section differ from a Background section?
You may remember that a Background section introduces and explains fundamental concepts that are important in the study, with the goal to improve a reader's understanding of later parts.
Conversely, a Related Work section discusses and compares (very) similar studies to the paper at hand.
Its goal is to show how this study is new or different and how it connects to and extends previous work.
It can be placed after the introduction, \eg in a combined Background \& Related Work section, or before the Conclusion section.
The first may be suitable if the need for the paper could be questioned by reviewers and should be strengthened early on, while the latter can be suitable if the need for the paper seems uncontested.
The advantage of presenting related work towards the end is that the study and its results are typically much more interesting for the reader.
Regardless, JB always places the Related Work section after the Introduction, usually in a combined Background \& Related Work section, with related work being covered in the last subsection.
RV only places the Related Work section after the Introduction if the novelty of the study has to be defended, otherwise before the Conclusion~section.

\subsubsection{Relate the Work}
It is very important for the quality of a related work discussion that this section is not treated as \enquote{annoying homework} to simply show that you know of other work in this area. 
Instead, the section should present an explicit comparison, \ie the main commonalities and differences, of your study and the related work.
For (nearly) every mentioned reference presented in this section, the relation to your study should be clearly described. 
For example, \enquote{Foo et al. present X. We also use~X, but apply it to a different context} or \enquote{Differently from the work of Foo et al., we use a different dataset to\dots}. 
In some cases, it is possible to group similar related work in a single paragraph and to summarize the common differences in it, \eg \enquote{The work of Foo et al. and Bar et al. use X on open-source software. In contrast, this work considers an industrial~context}.
In cases where a lot of similar related work already exists, it can also be helpful to create a table that makes differences in various categories easily identifiable.
You can find an example in the work of \citet{Martinez-Fernandez2022} and \citet{verdecchia2021building}.

\subsubsection{Use a Cohesive Logical Flow}
The second most important advice on writing the Related Work section is to structure its content to follow a logical flow that cohesively guides the reader.
This process starts by identifying the main topics covered by the related work, grouping the literature accordingly, and ensuring that a cohesive narrative binds all paragraphs together.
The number of papers to be discussed and the degree of relation to the presented study highly varies according to the research topic.
In a well-established research area, the related work might present a high degree of similarity that should be carefully discussed to defend the novelty and significance of the work.
In this case, the focus of the related work section would be narrow, by discussing in detail only the studies most closely related to the presented work.
However, sometimes, it might be difficult to find studies that are very similar to your own.
In this case, you have to expand the scope of the discussed related work, potentially in several directions.
For example, imagine you are the first to evaluate the impact of microservices patterns on maintainability through a controlled experiment with human participants.
Which related work should you discuss if nobody has done this before?
As a start, you could discuss studies evaluating the maintainability impact of patterns from other domains, \eg the Gang of Four design patterns.
One obvious next choice could be studies about the impact of microservices patterns on other quality attributes, \eg performance efficiency or reliability.
Lastly, you could also refer to some other studies about microservices maintainability or controlled experiments with microservices, even if they are not about patterns.

\subsubsection{Be Objective and Professional}
While discussing the comparison with related work, it is crucial to use a polite and objective stance.
Stay as truthful and fair to the content of the discussed studies as possible, compare concrete numbers whenever possible, and avoid presenting the work of your peers in an excessively bad light.
That does not mean that you are not allowed to point out what deserves criticism, but keep it professional and objective.

It is important that you gain thorough familiarity with the related work \textbf{before} the start of an ESE study. This ensures the effort invested in the research is justified, by presenting a novel contribution that is well-positioned within the existing state~of~the~art.

\textbf{Length:} Usually 1-2 columns.

\subsection{The Approach}
\label{sec:approach}
Not every ESE paper will provide a design contribution.
For example, the contribution of a typical controlled experiment or interview study is empirical evidence.
However, if the main contribution of your paper is a design contribution, then you will need a section that presents it.
Since many ESE papers provide (tool-supported) approaches as their design contributions, we call this section \enquote{Approach}.
However, depending on the nature of the contribution, other names are possible.
If your approach or tool has a concrete name, you can also include it into the section heading.
The goal of this section is to let the reader understand the nature of the provided design contribution and how it was created in the clearest, most unambiguous, and simplest way.
An example of an ESE paper with a design contribution called \texttt{RESTRuler} is presented by \citet{bogner_restruler_2024}.
In this paper, the respective section is called \enquote{Tool-Supported Approach (RESTRuler)}.
Additionally, the paper has a small \enquote{Research Process} section before that, which describes how \texttt{RESTRuler} was created.

By following Occam's Razor\footnote{\url{https://www.britannica.com/topic/Occams-razor}. Accessed 2025-03-20.}, the approach should be presented in the simplest way possible.
To this end, it is crucial to carefully understand the right abstraction level to present it by identifying the minimum amount of information necessary to understand the approach.
Implementation details, such as code snippets, comprehensive software design documentation, installation steps, \etc should be avoided as much as possible, as they can be provided in a digital appendix. 
Instead, strategically selected figures and / or pseudocode algorithms can be used to support the illustration of the approach in an intuitive and compact manner or to exemplarily describe some important parts of it in more details.

If the approach is characterized by different phases and / or steps, the Approach section can be subdivided into different subsections or paragraphs, by following the natural subdivision of the approach.
To ease the writing of this section and for a quick visual overview, a single figure can be created to summarize the approach holistically, and then the section is structured by simply describing the individual parts of the figure.
As a good presentation practice, the input and output of each step should be documented, respectively at the beginning or end of each subsection and potentially also in the figure, if space allows.

If the approach uses non-trivial processes or concepts from related work, it is recommended to describe these parts briefly in this section, rather than only relying on a reference without any summary.
This allows the reader to swiftly understand the approach in a self-contained manner, without the need to jump between references to gain a complete understanding of the approach.

\textbf{Length:} May highly vary, but usually 3-4 columns or less.

\begin{figure*}[ht!]
    \centering
    \includegraphics[width=\textwidth]{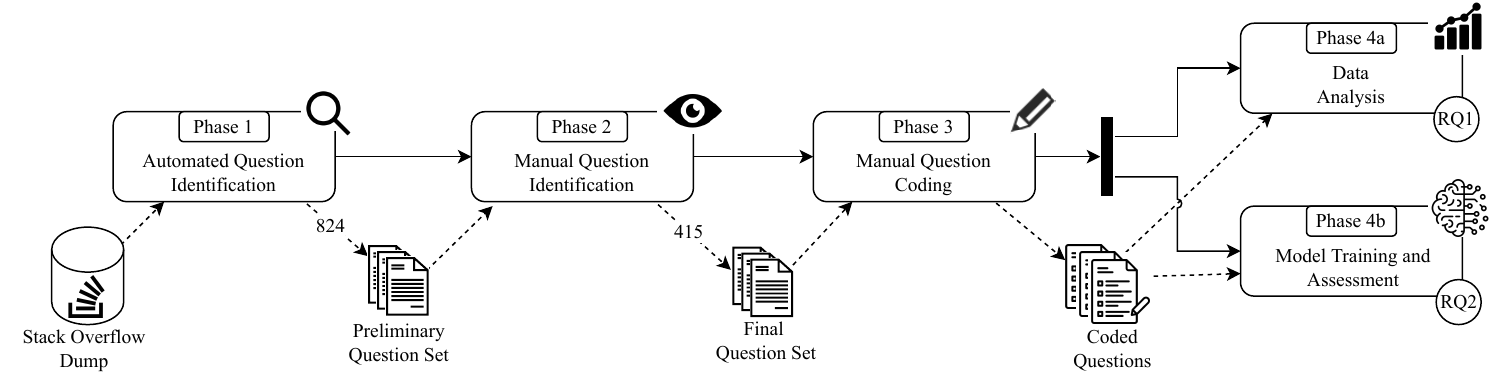}
    \caption{Example of a research process overview figure (taken from \citet{kozanidis2022asking})}
    \label{fig:process} 
\end{figure*}

\subsection{The Study Design}
The Study Design section provides an overview of the research process used to collect the results, followed by its step-by-step documentation.
If the paper centers around a design contribution and therefore has an Approach section, this section is typically called \enquote{Evaluation Design}, as the conducted study then typically is about evaluating the design contribution for its effectiveness (see again \citet{bogner_restruler_2024} for an example).
Similar as for the Approach section, selecting the correct level of detail is crucial, \eg low-level implementation details are omitted and made available for scrutiny in the replication package.
The main purpose of this section is to convince your reader that you followed a rigorous process that allowed you to arrive at valid results and conclusions, \ie it answers why they should believe you.
As its second purpose, this section should also allow other researchers to reproduce and replicate your results, in combination with an externally provided digital replication package.

\subsubsection{Research Objectives and Research Questions}
The Study Design section frequently opens with the formulation of the research objective, commonly by following the goal-question-metric (GQM) template~\cite{Basili94}, with the template portions highlighted with bold or italic font.
You can find examples in the work of \citet{Bogner2021a}, \citet{castano_exploring_2023}, and \citet{verdecchia2022tales}.
Note that it is not required to use such a structured approach to define your research objective.
While it can be helpful in the beginning when defining your objective, using GQM wrongly or inconsistently in your paper will instead anger your reviewers.
So, do not feel obligated to include it only because others do so.

After a potential research objective, the RQs of the study are reported.
As good practice, each RQ is usually supported by a paragraph that further details its motivation, the high-level benefits gained by answering the RQ, and any other information necessary to fully understand the RQ.
Each RQ should be a single, well-defined, self-contained question.
We recommend avoiding \enquote{yes / no} questions and converting them into \enquote{how} questions, e.g., \enquote{Does X influence Y?} $\Rightarrow$ \enquote{How does X infuence Y?}
As a ballpark figure, ESE papers usually have two to three and seldom more than five RQs.
For a more in-depth discussion of formulating research objectives and questions, we refer to \citet{Wieringa2014a}.

\subsubsection{Documenting Your Research Process}
Regarding the research process documentation, this description should enable the reader to understand, with sufficient details but without going down to the low-level implementation, how your study was designed, how the experimental data was collected and analyzed, and how conclusions were drawn.
Ensure you provide sufficient details for reviewers to assess the rigor and soundness of your study.
Failing to disclose important parts of your study design will typically be held against you by reviewers, without giving you the benefit of the doubt.
For the most important details, you should interleave the \textit{what} with the \textit{why}, \ie describe both what you did and why you did it (your rationale).
Alternate between the two, without only focusing on one of them.
Some decent examples where we tried to justify the major study design decisions include the papers of \citet{bogner_type_2022}, \citet{bogner_restful_2023}, \citet{migliorini2024architectural}, and \citet{maggi2025evolution}.

Before going into details, make sure to clearly state which research method(s) you have used.
Introduce each method with a solid reference, ideally an SE-related one\footnote{You can find a selection at \url{https://xjreb.github.io/swe-research-methods}.}, and briefly explain the method in a few sentences, potentially with pros and cons.
If you deviate from an established method, you need to explain why and provide convincing justification for this.
Moreover, never write that you used a certain research method, \eg grounded theory~\cite{Stol2016}, when you actually did \textit{not} do so.

If the process entails a complex structure, \eg multiple steps, phases, or intermediate inputs, a figure outlining the overall research process should be used to allow for a quick, high-level understanding of the study.
The section will then be developed starting from this visualization, with an opening paragraph describing the process overview, and each following subsection dedicated to describing a specific phase / step (see also Section~\ref{sec:approach}).
An example of such a research process visualization from a Stack Overflow mining study by \citet{kozanidis2022asking} is reported in Fig.~\ref{fig:process}, with another example in \citet{bogner_restruler_2024}.

Study designs are usually written in either the present tense -- to give a timeless, repeatable character -- or the past tense, since everything will be over when people read the paper.
Choose one of the two and consistently stick to it.

Commonly, the study design of ESE papers broadly covers three main phases:
\begin{itemize}
    \item Phase 1: Identification and Pre-processing of Experimental Subjects or Objects (\eg repository or participant selection)
    \item Phase 2: Data Collection (\eg experiment measurements, interview transcription, or source code metric collection)
    \item Phase 3: Data Analysis and Synthesis (\eg data labeling or statistical analysis)
\end{itemize}

Usually, each phase is characterized by one or more steps, with the three main phases often left implicit to highlight the nature of the single steps better.
The length of each phase may vary depending on the used research method(s).
For example, in a case study, the identification phase might be completely omitted, as related information might already be provided in the Introduction.

\subsubsection{Phase 1: Experimental Subject / Object Identification} This part is also called \textit{sampling}~\cite{Baltes2022} and describes the process leading to the systematic identification and selection of your experimental subjects or objects, \eg a set of software repositories, interview participants, online data sources, or competitor approaches.
This documentation should thoroughly describe the selection process, supported by a clear rationale and potentially the objective selection criteria used for their identification.
You need to make it effortless to clearly understand why your selection is relevant and representative for answering the RQs of your study.

\subsubsection{Phase 2: Data collection} This part provides a detailed description of how the study data was collected.
Data collection highly varies based on the used research method, and can range from measurements during a controlled experiment, to field notes during an ethnographic study, or the metric-based analysis of source code repositories.
In some cases, it can also make sense to split this section into a Study Execution and a Data Collection section.
The first covers how the different steps of the study were operationalized, plus information on execution details like study duration, while the latter covers data collection in greater detail.
This can be especially helpful if the data collection is very extensive~and~complex.

If the process uses a set of dependent and independent variables defined \textit{a priori}, these variables can be documented in one or two dedicated subsections.
All variables and their operationalization, \ie how you will measure them, should be defined with the utmost care, and supported by a clear and transparent rationale.
Similarly, if data is extracted \textit{via a priori} defined criteria, \eg through provisional coding, this section can document the definition of the criteria used by the researchers.

The data collection steps should be documented in a clear, self-contained fashion, and ideally be supported by references.
On one hand, the data collection process should communicate to the reader \textit{what} was done in sufficient detail.
On the other hand, it is important to also communicate \textit{how} and \textit{why} this process was followed, to demonstrate that the data collection was systematic and reliable, without major impact of potential confounders.
Lastly, do not forget to also describe the environment in which the data was collected.
For example, if measuring software performance via dynamic analysis is part of the data collection, you also have to provide information regarding the hardware used.
Similarly, you would have to describe the environment in which human subjects participated in a controlled experiment or interview study.
With human participants, describing ethical aspects, \eg how you obtained informed consent, is especially important.

\subsubsection{Phase 3: Data analysis} This part documents how the collected data is analyzed.
Commonly, this portion reports the labeling processes and / or statistical analyses used to derive insights from the data, evaluate hypotheses, and answer RQs.
Initial supporting steps like data harmonization, outlier-removal, and the creation of various plots are also reported in this section.

\textbf{Length:} Usually around 3-5 columns for the whole section.

\subsection{The Results}
In the Results section, the final results originating from the data analysis and synthesis are reported.
While presenting the results, ESE authors usually refrain from also interpreting the results, as this latter part constitutes the content of the Discussion section.
Results should therefore be presented as objectively as possible, \eg in case of a quantitative analysis, just by describing the obtained measurements and results of the hypothesis testing.
Graphs, tables, and other visual aids, \eg snippets, should be used to support the textual presentation of the results (for tips on figures, see Section~\ref{sec:figures}).

To make the presentation of the results more relatable and digestible, specific examples can be used to illustrate a particular result in an anecdotal but concrete manner, \eg an interview participant quote, a code snippet, or a specific measurement.

Commonly, results are presented by starting with initial, general findings (if they exist) and then dedicating each subsection to answer a RQ or specific identified topic, \eg in case of a qualitative-first research.
If you use one subsection per RQ, you ideally should not simply use \enquote{5.1 RQ1} as the heading and also do not use the complete, potentially long RQ.
Instead, paraphrase the RQ into a concise but thematically fitting heading.
If you like, you can add the respective RQ at the end in parentheses.
You can see examples of this in \citet{Bogner2021a} and \citet{bogner_restful_2023}.

Within each subsection, results should also follow a systematic and organized style, possibly by considering sub-RQs and a paragraph or subsection order that all results share, \eg a summary results overview, descriptive statistics with figures, hypothesis testing results, concrete examples, and summary takeaways.

While writing the results, try to provide concrete numbers whenever possible and highlight the most important aspects of the results.
You should strive to be as precise and comprehensive as possible, while also focusing on what is really important to report.
Especially if your results are very extensive and rich, it is good practice to provide boxed environments at the end of each subsection to highlight the main findings for the reader, \eg as concise answers to the RQs.
Boxed environments can also or alternatively be used to highlight \enquote{findings}, which are not necessarily mapped to a specific RQ.
This finding text should be as concise and as descriptive as possible, summarizing the main findings in the most succinct and expressive way possible.
Whenever possible, concrete numbers should be reported, \eg \enquote{10\% accuracy improvement over the competitors}.
A dummy example of an RQ answer highlight is shown below.
For concrete examples, we refer to \citet{maggi2024claim} or \citet{Bogner2021}.
Alternatively, you can also report (or briefly repeat) the main answers to the RQs at the beginning of the Discussion section.
RV does this frequently, while JB mostly refrains from doing so.

\begin{highlight}
    \faIcon{lightbulb} \textbf{Answer to $\mathbf{RQ_1}$ (Highlights):} Boxed environments can be used to summarize RQ answers or main findings for readers who just wish to skim through the paper.
\end{highlight}

\textbf{Length:} Usually around 6-8 columns

\subsection{The Discussion}
\label{sec:discussion}
While the Results section focuses on a neutral and objective presentation of your findings, the Discussion section provides a slightly more subjective discussion of the interpretation (\enquote{why are some results the way they are?}) and implications (\enquote{what should we do now?}).
Phrasings such as \enquote{we conjecture}, \enquote{our hypothesis to explain this is}, \enquote{this seems to be due to}, and other subjective phrases may appear more often in this section than in others.
In other words, while the Results section is intended for a reader who is interested in the \enquote{raw}, unprocessed, and objective results, the Discussion section takes this further by providing the reader an explanation of the results, what they mean, and their potential implications, \eg for researchers, practitioners, educators, \etc
Try to focus as much as possible on \textit{direct} implications of the results and do not exclusively mention future research that is required for further insights.
It is possible to mention important identified research gaps (especially in secondary studies), but future work should typically be proposed in the Conclusion section.

Discussing the results like this also allows providing further details on the findings, \eg by discussing cherry-picked examples, corner cases, presenting cross-RQ result analyses, and discussing the relation of the results with related work.
The Discussion section is typically organized into subsections, with each subsection dedicated to discussing the findings of a specific RQ.

In addition to the interpretation and implications of the RQ results, the Discussion section can also be used to report other interesting observations, lessons learned, recommendations, or takeaways that do not necessarily have a very strong connection to the RQs.
In line with the nature of the Discussion section, this content expands the findings to provide a bigger picture, \eg what have we learned from this study? or what do the results mean for research and / or practice?
Try to focus as much as possible on direct implications of the result and do not exclusively talk about future research that is required for further insights. It's possible to mention important identified research gaps (especially in secondary studies), but future work should typically be proposed in the conclusion section.

In case of a tight page limit or if the Results section is very long, the Discussion can also be merged with the Results section (referred to as either just \enquote{Results} or \enquote{Results \& Discussion} section).
In this case, parts of the results are typically first presented and then discussed, by following the order provided by the RQs.
More specifically, a subsection could be dedicated to each RQ, composed of the RQ results presentation, the discussion of the results, and optionally a takeaway summary.
You can find an example of such an intertwined Results and Discussion section in \citet{bogner_restful_2023}.

\textbf{Length:} Usually around 1-3 columns.

\subsection{The Threats to Validity}
This section is used to openly and transparently document potential threats that may have influenced the validity of the results.
The content of this section is paramount to let the reader comprehensively understand and judge the results.
Ideally, threats are documented by following a threat categorization that fits the study, \eg the ones presented by \citet{wohlin_experimentation_2024} in case of controlled experiments.
A dedicated subsection or paragraph should be used for each threat category, in which you document the most important or most likely threats that may have influenced the results.
Each threat, in addition to its description, should be supported by how you tried to mitigate it or a description of why using a mitigation strategy was impossible.
However, do not only describe threats that you mitigated: also discuss which threats remain and how likely it is that they impact your results.
If possible, provide existing evidence for how strongly potential threats may (or may not) have influenced the results (see \citet{Wyrich2024} for inspiration).

Ideally, you should distinguish between \textit{limitations} and \textit{threats to validity}.
Limitations are usually conscious decisions that reduce the scope of your approach to make it more practical or effective, \eg your prototypical tool support might only focus on the analysis of Java-based systems.
Threats to validity, on the other hand, are usually not conscious decisions, but issues that are difficult to mitigate completely, such as subjective researcher bias in qualitative research.
However, you can also describe some threats to validity as conscious trade-offs, \eg sacrificing external validity for increased internal validity by sampling a more homogeneous set of participants.
As an alternative to describing such conscious decisions in this section, they can also be described in the Study Design section, leaving the Threats to Validity section for mostly unaddressed and late-emerging threats.
Other common pitfalls to avoid while considering and documenting threats to validity are described by \citet{verdecchia2023threats}.

\textbf{Length:} Usually around 1-1.5 columns.

\subsection{The Conclusion}
This section closes a paper by reporting a very short summary of the paper content (usually 2-3 sentences), followed by what can be \textit{concluded} from the results.
The Conclusion (also plural \enquote{Conclusions}) makes a step beyond the Discussion, by conveying the main message and lesson learned from the study.
Your conclusions should not be a rehash of the results, but rather a thoughtful and concise description of the essence of the entire paper, from its initial motivation to the final results and their discussion.
In other words, this section gives a closing overview of why the study is important, what was learned from the study, and how others can profit from~it.

The Conclusion usually ends with 1-2 paragraphs on future work.
Such content is used to describe for the research community how the authors intend to build upon the study or how they plan to mitigate limitations, to inspire others to extend the work, and to encourage future collaborations on the topic. 

\textbf{Length:} Usually around 0.5-0.75 columns.

\subsection{The Acknowledgments}
Typically, the Acknowledgments section is not required unless the research has benefited from contributions made by individuals besides the authors or specific funding schemes.
Acknowledgments are commonly formulated as one or two sentences.
Usually, the name of the contributor followed by the type of contribution is documented, \eg \enquote{We express our sincere gratitude to X for their insightful review of the paper draft.}
Research grant acknowledgments typically follow a template provided by the funding agency.

\textbf{Length:} Usually around 3-4 lines.

\subsection{Open Science and Replication Packages}
In the interest of Open Science~\cite{Mendez2020}, it is important to provide your study artifacts as a publicly available digital appendix that is linked in the paper.
Instead of using something like Google Drive or a GitHub repository, you should use a service that provides a digital object identifier (DOI) and is committed to long-term archiving.
Good options are, \eg Zenodo\footnote{\url{https://zenodo.org}} or Figshare\footnote{\url{https://figshare.com}}, but some institutional repositories also provide this.
If you have reusable software contributions like developed tool support, using a GitHub repository is still a good option for usability and discoverability, but make sure to also archive it on Zenodo to get a DOI.\footnote{\url{https://docs.github.com/en/repositories/archiving-a-github-repository/referencing-and-citing-content}}
Your replication package should include, \eg your data (study results), code (experiment execution scripts, data analysis scripts, \etc), or other artifacts that you created during the study and that are valuable to increase transparency, reproducibility, and reusability.
Additionally, there should be at least a minimum of documentation that explains the artifacts and how to use them.
You can find a template for such a replication package in the S2 GitHub organization.\footnote{\url{https://github.com/S2-group/template-replication-package}}

Reusing research artifacts should be simple, \eg for Python code, provide a \texttt{virtual env} or a \texttt{requirements.txt} and the commands to quickly set up and use everything.
A Docker image\footnote{\url{https://docs.docker.com}} can also be a convenient alternative for an otherwise complicated setup.
Lastly, be careful to only share artifacts that you are allowed to share publicly, \eg private data of interviewees needs to be protected.
Ensure that you obtain informed consent from your study participants regarding data sharing.

\subsection{Section Writing Order}
\label{sec:writing_order}
While some sections like the Related Work, Approach, and Study Design can be written independently, others need to be addressed towards the beginning or end of the writing process.
Therefore, writing a paper linearly from Abstract to Conclusion is usually not the best idea.
A potential section writing order, dictated solely by the content of the sections, could be the following.\footnote{Note: the section writing order may vary according to the specific research methods or personal preference.}

\begin{enumerate}
    \item Results [could be swapped with Study Design, but must precede the Discussion]
    \item Discussion [ideally written just after the Results]
    \item Study Design [could also be written first]
    \item Approach [order-independent]
    \item Threats to Validity [order-independent]
    \item Background [order-independent]
    \item Related Work [order-independent, but some people like to write it early, even before thinking about the study design]
    \item Introduction [mostly order-independent, but ideally written towards the end when the paper contributions are stable]
    \item Conclusion [ideally written towards the end]
    \item Abstract [ideally written last]
\end{enumerate} 

\section{General Scientific Writing Tips}
In the following sections, we provide a collection of more general principles and tips for scientific writing.
These guidelines are mostly based on seminars we attended or advice we received from more experienced writers.
They may be useful beyond ESE papers, \eg for (computer) science research papers in general.
However, keep in mind that these are mostly \textit{principles}, not \textit{rules}, meaning there can be valid reasons to occasionally violate them because another concern is more important.

\subsection{Concreteness Over Abstraction}
You are a researcher, not a politician.
Therefore, it is important that your statements are concrete and precise.
For example, consider the following sentence:
\enquote{We relied on rigorous data cleaning procedures.}
What did you do exactly?
Without more concrete details, this can mean very different things.
Similarly, avoid both excessively sounding adjectives and adverbs (\enquote{enormous}, \enquote{gigantic}, \etc) and vague or subjective words (\enquote{some}, \enquote{easy}, \enquote{fast}, \etc).
Ideally, try to provide numbers in such cases.
Furthermore, avoid making unsubstantiated bold claims.
The stronger your claim, the more important it is to provide credible references for it.
You can get away with not providing a reference for a claim that has broad consensus in the community and is formulated in a fairly neutral way.
For everything else, provide a solid reference.
Lastly, in the Conclusion, take responsibility for your research and formulate concrete take-home messages.
However, also be humble when applicable and know the limitations of your research.

\subsection{Avoid Wordiness, Be Concise}
You may have heard the following quote from Blaise Pascal:
\enquote{I would have written a shorter letter, but I did not have the time.}\footnote{\url{https://en.wikipedia.org/wiki/Lettres_provinciales}}
Writing concise papers that focus on the important things is difficult, but it is something we should all strive for.
Therefore, if you can omit words without substantially changing the meaning, you should usually do so.
In the same way, avoid very long sentences.
As a test, read such a sentence aloud.
If you struggle to breathe or lose the meaning of it, the sentence should be shortened and / or split up.
A simple example of unnecessary wordiness is the use of \enquote{in order to}.
Something like \enquote{In order to reach our goal, ...} can be changed into \enquote{To reach our goal, ...} without losing anything.
Similarly, avoid writing sentences such as \enquote{It is important to note that ...}. 
Usually, such wordy parts of a sentence can~simply~be~removed.

\textbf{Do not repeat yourself.}
Each concept should be presented exactly once throughout the entire document.
Ideally, it should be positioned where a reader is most likely to look for it.
If a previously presented concept needs to be reintroduced, use cross-references instead of repeating the concept. 
Give your new approach or tool a short name.
This makes the contribution easy to remember and eases the writing, \eg you will not have to repeat \enquote{the approach proposed in this paper} over and over.
Ideally, the name should be a meaningful and memorable acronym, \eg FAST Approaches to Similarity-based Testing (FAST)~\cite{miranda2018fast}.

\textbf{Use the right level of detail.}
Provide sufficient information to allow readers to verify the soundness of your study design, but remember that some low-level details cannot be included for space reasons.
Additionally, avoid documenting a research travelogue.
Report only what is important to understand the final approach and research process (an exception may be Design Science Research papers~\cite{hevner_three_2007}, which follow cycles by nature).
Do not include all details about preliminary experimentation, failed attempts, unused variations of a new algorithm, \etc
Lastly, for the results, also make sure that the main messages or answers to the RQs are not lost in extensive details.

\subsection{Use an Active and Simple Writing Style}

\textbf{Active over passive.}
Favor active and dynamic sentences, and do not be afraid to use first person (\enquote{we}).
Instead of excessively using the passive voice, favor \enquote{flesh-in-blood} subjects positioned closely to the verb.
Using active voice makes sentences short, direct, engaging, and easy to understand, \eg consider \enquote{It has to be highlighted by us that ...} vs. \enquote{We highlight that ...}.
Similarly, avoid \enquote{zombie nouns} as fake subjects, \eg \enquote{A paper by Verdecchia and Bogner describes ...}.
A paper cannot describe anything, the \textit{authors} do that.
Should you never use passive voice then?
No.
Sometimes, passive voice can be consciously and sparingly (!) used to indicate distance, generalizability, and repeatability, \eg in the Study Design section.

\textbf{Try to keep everything simple.}
Contrary to what some people seem to believe, there is no need to impress your readers with complex language in research papers.
Therefore, prefer simple yet non-trivial language as much as possible.
Use technical or more sophisticated terms for additional precision if necessary, but \textit{not} to sound more knowledgeable.
Be warned that text produced or \enquote{improved} by tools like ChatGPT is often guilty of this.
Typical red flags are words like \enquote{pertain} or \enquote{utilize}, \eg \enquote{Pertaining to service deployment, we utilized a sophisticated script-based mechanism.}
The same information can be expressed as \enquote{We used a shell script for service deployment.}
Similarly, avoid using rare, esoteric words like, \eg \enquote{adjudged}.
If you are not a native English speaker, do not look up single-word translations and then pick a word you never heard before just because it sounds more fancy.
It may be an odd or even incorrect choice in your specific context.
Keeping your language and explanations simple is also important because your reviewers may not always be experts on the topic or research method.
In the end, researchers will like your paper if they have the feeling that they understand it, so make it easy for them.

\subsection{Be Consistent and Conscientious}
You are not only a researcher, but also a computer scientist / software engineer.
Consistency is very important for research, but also in our professional discipline in general.
Therefore, name and write the same things in exactly the same way.
This includes, \eg important concepts, capitalization, referencing, formatting, usage of spaces, headings, colors, figures, and tables.
How can reviewers trust you that you have been consistent throughout your study if you are not consistent within a paper you write?
This consistency should extend to your paper structure.
All content should be located in its appropriate section or subsection.
Whenever possible, reuse the same paragraph order, \eg for each RQ, you could always present the raw results, data analysis, and then the interpretation.
The same is true for reusing a potential order of bullet points in similar listings.
Each section should also consistently use the same main tense, \eg do not mix simple past and present in the Study Design.
Lastly, try to be consistent in keeping all paragraphs a similar length.
They should neither consist of only one or two sentences, nor should they be a \enquote{wall of text} that merges dozens of sentences in a single paragraph.

\textbf{Be conscientious.}
Everybody makes mistakes, so double-check what you wrote.
Ensure that your chosen phrasing or structure really works well.
Ensure that you did not make any random typos or accidentally omitted a word.
Ensure that you provide evidence for all major claims.
Ensure that you always provide the same numbers for the same concepts throughout the whole paper, \eg in systematic literature reviews (SLRs).
Ensure that you use proper punctuation.
Ensure that you use consistent capitalization for section names and captions.
Proofread what you wrote before you send it out to others for review.\footnote{\textbf{Important caveat:} do not paralyze yourself with perfectionism. A first complete draft does not have to be perfect. It just needs to be of sufficiently high quality that your co-authors can easily understand it. This allows them to fully focus on the high-value feedback: content, not presentation.}
Unfortunately, conscientiousness is difficult to learn because it is a fairly stable character trait (see, \eg the Big Five personality traits / OCEAN model\footnote{\url{https://en.wikipedia.org/wiki/Big_Five_personality_traits}.}).
Therefore, remind yourself about double-checking and use tool support, such as LanguageTool\footnote{\url{https://languagetool.org}}, Grammarly\footnote{\url{https://www.grammarly.com}}, DeepL Write\footnote{\url{https://www.deepl.com/en/write}}, or Hemmingway App\footnote{\url{https://hemingwayapp.com}}.
Similarly, avoid improvising the content of sections, subsections, and paragraphs as you go, especially if you are fairly inexperienced with scientific writing.
Think about the major points or the chain of arguments you want to make, \eg in a subsection.
If it helps you, you can also write down a short bullet point for each argument and add supporting references.
Once you are happy with the general flow, convert each bullet point into a paragraph.

\subsection{Know and Guide Your Reader}
This principle is important for all types of writing, not just scientific writing.
When you write an ESE paper, the first external people reading your manuscript are the reviewers.
Reviewers are typically researchers themselves, \ie the final audience for your paper.
Ask yourself: what do they expect?
What are they used to?
What could make it easier for them to review the paper?
Under which circumstances do they sometimes review papers?
How can you suitably manage their expectations?

\textbf{Guide the reader through the content of the paper.}
The reader should never feel puzzled about why they are reading a certain section or paragraph.
At the macro level, sections and sometimes even longer subsections can be briefly introduced right after the section header to provide a brief guiding overview of the content to come.
Avoid stacking several section and subsection headers right after each other, especially more than two.
At a more fine-grained level, paragraphs should be logically linked, without abruptly jumping from one topic to the next without a logical link.
The motivation for going from one paragraph to the next should feel intuitive.
Construct a consistent reading flow, with a clear narrative throughout the paper.
After finishing a paragraph, read it again a couple of times.
Make sure you always know where you are, both at the micro-sentence level and macro-paper level.
If you get lost, the sentence or flow is likely not clear or too complex.
Similarly, avoid using parentheses to structure sentences, as it frequently leads to overly complex structures that break the natural reading flow.
Overall, try to make the content browsable.
A reader should be able to find the information or parts they are looking for in the paper without much effort.

\section{Language, Formatting, and \LaTeX~Tips}
Lastly, in the remainder of this section, we compiled a collection of smaller, practical tips and guidelines related to scientific writing, formatting, and essential \LaTeX best practices. 

\subsection{Language \& Text Formatting}
The following list represents some basic tips regarding language and text formatting.

\begin{itemize}
    \item Use capitalization, italics, bold, and other special formatting \textit{sparingly}. For example, capitalization should only be used at the beginning of sentences, to write names, and to introduce acronyms, but even the latter is not strictly necessary.
    \item Numbers below 10 are usually written in letters, \eg \enquote{nine projects}. However, many style guides also tell you to never start a sentence with a number written with digits, \eg \enquote{15 people attended.} $\Rightarrow$ \enquote{Fifteen people attended.} or \enquote{In the end, 15 people attended.} Similarly, some style guides tell you to never mix numbers written as words with numbers written as digits in the same sentence, \eg \enquote{We had 12 blue ones and four red ones.} $\Rightarrow$ \enquote{We had 12 blue ones and 4 red ones.}
    \item In English, the decimal sign for numbers is the dot, \eg 3.14159, while the delimiter sign is the comma, \eg 100,000. Use the latter for all numbers greater than 999.
    \item Do not use contractions, \eg \enquote{can't} $\Rightarrow$ \enquote{cannot}.
    \item \enquote{\ie} and \enquote{\eg} should always be followed by a comma.
    \item Sentences starting with \enquote{\eg} should not end with \enquote{\etc} because \enquote{\eg} means \enquote{for example}, while \enquote{\etc} is used to be more comprehensive.
    \item There is a difference between \textit{dependent} and \textit{independent} relative clauses regarding comma usage. Do not use a comma for a dependent relative clause, \eg \enquote{The man who stole my money was bald.} However, for independent relative clauses, you must use a comma, \eg \enquote{I won’t get my money back, which is not a big deal.} Omitting a comma here would change the semantics. This is also why you never use a comma before \enquote{that} and \enquote{because}: they always indicate a dependent relative clause. The only exception is if the dependent clause comes \textit{before} the main clause: \enquote{I will follow wherever you go.} $\Rightarrow$ \enquote{Wherever you go, I will follow.}
    \item Always use a comma after an introductory word or expression, \eg \enquote{Furthermore, \dots} or \enquote{After the match, \dots}.
    \item We suggest using the Oxford comma\footnote{\url{https://en.wikipedia.org/wiki/Serial_comma}. Accessed 2025-03-20.} because not doing so can cause ambiguity. Take a look at this example: \enquote{We invited the students, Roberto, and Justus.} Here it is clear: we invite the students plus two more. However, without the Oxford comma, it becomes ambiguous: \enquote{We invited the students, Roberto and Justus.} Are Roberto and Justus the two invited students? It is not fully clear.
    \item Clarify with your coauthors if you write the paper in British or American English, and consistently stick to this, \eg \enquote{analyse} vs. \enquote{analyze}, \enquote{behaviour} vs. \enquote{behavior}, or \enquote{modelling} vs. \enquote{modeling}.
    \item Always use the word \enquote{Section X}, even when referring to subsections, subsubsections, \etc
    \item Avoid using subsequent words that sound similar, \eg \enquote{we \textit{extract} and \textit{abstract} from \dots}
\end{itemize}

\subsection{\LaTeX~Tips}
The following list covers some basic \LaTeX~formatting tips.
For the interested reader, a far more exhaustive list has been compiled by Spinellis.\footnote{\url{https://github.com/dspinellis/latex-advice}. Accessed 2025-03-20.}

\begin{itemize}
    \item If you are not writing a book or a long monographic dissertation, use a \textit{single} \texttt{.tex} file. This makes it easier to edit, navigate, find \& replace text, copy \& paste text, and browse from the \texttt{.tex} file to the PDF document and back. Following standard notation, the file should be called \texttt{main.tex}.
    \item Put every sentence in its own line. This allows precise navigation from the PDF view to the \LaTeX~view and also makes  diffing of two versions easier.
    \item Use a non-breaking space (\(\sim \)) before all \texttt{\textbackslash cite\{\}} and \texttt{\textbackslash ref\{\}} commands to avoid line-breaking.
    \item The \texttt{natbib} package\footnote{\url{https://ctan.org/pkg/natbib}} has convenient commands for different types of citations, e.g., \texttt{\textbackslash citeauthor\{\}} or \texttt{\textbackslash citet\{\}}. With these, you are never in danger of misspelling a complicated author name or falsely using \enquote{et al.} with two authors.
    \item Labels should be meaningful and be used with consistency. Section labels should be preceded by \texttt{sec:}, figure labels by \texttt{fig:}, and table labels by \texttt{tab:}. Whenever possible, labels should be the first one or two words of the section, figure, or table name (or an understandable contraction of it, \eg \texttt{\textbackslash sec:intro}). This makes labels easy to remember and~use. The same reasoning applies to file names, \eg call a figure \texttt{results-rq1-performance.pdf} instead of \texttt{figure3.pdf}.
    \item All URLs, references, and footnotes should be clickable, e.g., by including the \texttt{hyperref} package.\footnote{\url{https://ctan.org/pkg/hyperref}}
    \item Avoid formatting subtitles with the \texttt{\textbackslash subtitle\{\}} command, as it might lead to issues when the paper will be indexed by digital libraries. Instead, use a colon to transform the subtitle as part of the main title, \ie \textless title\textgreater: \textless subtitle\textgreater.
    \item If you start exceeding the page limit, some common shortening strategies are (i) shortening paragraphs with just a few words on a new line, if needed with the support of \(\sim \), (ii) reducing the size of figures, (iii) reducing the font size in tables, (iv) removing content deemed less important, and (v) editing references to remove redundant content, \eg removing a URL if a DOI is already there. We generally want to discourage using space-saving libraries such as \texttt{savetrees}\footnote{\url{https://ctan.org/pkg/savetrees}} or the \texttt{\textbackslash vspace\{\}} command. Such \enquote{space-saving hacks} might cause issues during the later editorial process or even get your paper rejected by some venues.
    \item If you find it helpful, use macros for the consistent formatting of repeated words and expressions, \eg the name of your tool or acronyms. However, keep this somewhat balanced, \ie avoid the excessive use of macros to shorten natural language words, \eg \textbackslash MSA for \enquote{Microservice-based Architecture}. While it speeds up the writing, the macros might make the \texttt{.tex} file harder to read and navigate for collaborators, as the natural flow of sentences deteriorates.
    \item Keep the \LaTeX~project clean. Old text should not be commented out but removed (make use of the history of Git or Overleaf for returning to old versions), unused files and figures should not be present, all figures should be placed in a dedicated folder, etc.
    \item Fix all \LaTeX~compilation errors immediately. If you postpone these and keep adding new things to the paper, it may become very difficult to fix the errors later. Regarding compiler warnings, at least inspect them regularly and fix those that are important to you (many can be safely ignored).
    \item Do not use \texttt{\textbackslash newline} or \texttt{\textbackslash\textbackslash} in titles and author blocks, as they may break the correct indexing of the paper metadata.
    \item When submitting a paper for review, some reviewers appreciate line numbers in the PDF. Most publisher templates have a special review mode that you can use for this.
    \item Use a consistent style to write BibTeX entry keys. For example, the one used by Google Scholar provides a good balance of semantic and syntactic information while making keys easy to link to the paper content (<first-author-name><year><first-title-word>). But even more important than consistency here is that the keys give at least some hint about which paper they refer to. Using DOIs or other random IDs is strongly discouraged.
\end{itemize}

\subsection{Figure Formatting}
\label{sec:figures}
Figures are one of the most noticeable portions of a paper.
As such, some readers might subconsciously start to judge the quality of the paper based on its figures.
In other words, figure quality is very important.
While the style of the figures greatly varies according to their content, some general good practice are:
\begin{itemize}
    \item Provide all figures in a vector-based format, \eg~\texttt{.pdf}, and ensure that no parts of the figure are pixelated.
    \item Avoid small text. As a rule of thumb, all text in the figure should be at least the size of the caption font. A figure should ideally be readable at 100\% zoom.
    \item The axes should be clearly labeled.
    \item In general, try to avoid using logarithmic scales or starting axes from something else than zero. However, there are definitely exceptions to this rule, \eg if readability would otherwise be strongly impacted.
    \item All text should be readable without having to tilt the head or paper. For example, do not incline text over \(\sim \)45$^{\circ}$. Similarly, do not rotate large tables or figures by 90$^{\circ}$. If you believe only this will ensure its readability, then this is a clear sign that you should fundamentally rethink and redesign it.
    \item If possible, each sub-figure should have its own caption.
    \item Reference all tables and figures in the text. They are there to support the understanding of the text. A non-referenced figure or table is a clear sign that it is currently not well-connected to the text. Additionally, ensure that figures and tables are not positioned too far away from the paragraph where they are described.
    \item Use a consistent visual theme for all plots in the paper.
    \item In figures like barplots, think systematically about the order of the plot elements. Which order would support understanding the data the most for your readers? Which questions might they have? Often, ordering by frequency will be most helpful, but exceptions are possible.
\end{itemize}

\section{Conclusion}
We realize this document has turned into a very dense collection of guidelines and tips, which might feel overwhelming to junior researchers.
If it does, do not worry!
You do not have to adhere to all of this at once.
Start by applying just a few of the guidelines that you find most reasonable, and then gradually include more once the initial ones start to feel natural to you.
You can also simply return to this document when you have trouble writing a specific section for a paper.
In the end, we encourage you to find your own \textit{style} of writing over the years.
There is more than one way of writing compelling ESE papers, so develop and train your personal scientific \enquote{voice} to communicate your research results.
We hope this paper can at least partially empower you to do that.
We wish you much success and fun with your writing!

\bibliographystyle{plainnat}
\bibliography{biblio}

\begin{thebibliography}{36}
\providecommand{\natexlab}[1]{#1}
\providecommand{\url}[1]{\texttt{#1}}
\expandafter\ifx\csname urlstyle\endcsname\relax
  \providecommand{\doi}[1]{doi: #1}\else
  \providecommand{\doi}{doi: \begingroup \urlstyle{rm}\Url}\fi

\bibitem[Arab et~al.(2022)Arab, LaToza, Liang, and Ko]{arab2022exploratory}
Maryam Arab, Thomas~D LaToza, Jenny Liang, and Amy~J Ko.
\newblock An exploratory study of sharing strategic programming knowledge.
\newblock In \emph{Proceedings of the 2022 CHI Conference on Human Factors in Computing Systems}, pages 1--15, 2022.

\bibitem[Baltes and Ralph(2022)]{Baltes2022}
Sebastian Baltes and Paul Ralph.
\newblock Sampling in software engineering research: A critical review and guidelines.
\newblock \emph{Empirical Software Engineering}, 27\penalty0 (4):\penalty0 94, July 2022.
\newblock ISSN 1382-3256.
\newblock \doi{10.1007/s10664-021-10072-8}.

\bibitem[Basili et~al.(1994)Basili, Caldiera, and Rombach]{Basili94}
Victor~R. Basili, Gianluigi Caldiera, and Dieter~H. Rombach.
\newblock \emph{{T}he {G}oal {Q}uestion {M}etric {A}pproach}, volume~I.
\newblock John Wiley \& Sons, 1994.

\bibitem[Bogner and Merkel(2022)]{bogner_type_2022}
Justus Bogner and Manuel Merkel.
\newblock To {Type} or {Not} to {Type}? {A} {Systematic} {Comparison} of the {Software} {Quality} of {JavaScript} and {TypeScript} {Applications} on {GitHub}.
\newblock In \emph{2022 {IEEE}/{ACM} 19th {International} {Conference} on {Mining} {Software} {Repositories} ({MSR})}, pages 658--669, May 2022.
\newblock \doi{10.1145/3524842.3528454}.
\newblock ISSN: 2574-3864.

\bibitem[Bogner et~al.(2021{\natexlab{a}})Bogner, Fritzsch, Wagner, and Zimmermann]{Bogner2021a}
Justus Bogner, Jonas Fritzsch, Stefan Wagner, and Alfred Zimmermann.
\newblock Industry practices and challenges for the evolvability assurance of microservices.
\newblock \emph{Empirical Software Engineering}, 26\penalty0 (5):\penalty0 104, July 2021{\natexlab{a}}.
\newblock ISSN 1382-3256.
\newblock \doi{10.1007/s10664-021-09999-9}.
\newblock Publisher: Springer.

\bibitem[Bogner et~al.(2021{\natexlab{b}})Bogner, Verdecchia, and Gerostathopoulos]{Bogner2021}
Justus Bogner, Roberto Verdecchia, and Ilias Gerostathopoulos.
\newblock Characterizing {{Technical Debt}} and {{Antipatterns}} in {{AI-Based Systems}}: {{A Systematic Mapping Study}}.
\newblock In \emph{2021 {{IEEE}}/{{ACM International Conference}} on {{Technical Debt}} ({{TechDebt}})}, pages 64--73. IEEE, May 2021{\natexlab{b}}.
\newblock ISBN 978-1-66541-405-0.
\newblock \doi{10.1109/TechDebt52882.2021.00016}.

\bibitem[Bogner et~al.(2023)Bogner, Kotstein, and Pfaff]{bogner_restful_2023}
Justus Bogner, Sebastian Kotstein, and Timo Pfaff.
\newblock Do {RESTful} {API} design rules have an impact on the understandability of {Web} {APIs}?
\newblock \emph{Empirical Software Engineering}, 28\penalty0 (6):\penalty0 132, November 2023.
\newblock ISSN 1382-3256, 1573-7616.
\newblock \doi{10.1007/s10664-023-10367-y}.

\bibitem[Bogner et~al.(2024{\natexlab{a}})Bogner, Kotstein, Abajirov, Ernst, and Merkel]{bogner_restruler_2024}
Justus Bogner, Sebastian Kotstein, Daniel Abajirov, Timothy Ernst, and Manuel Merkel.
\newblock {RESTRuler}: {Towards} {Automatically} {Identifying} {Violations} of {RESTful} {Design} {Rules} in {Web} {APIs}.
\newblock In \emph{2024 {IEEE} 21st {International} {Conference} on {Software} {Architecture} ({ICSA})}, pages 123--134, Hyderabad, India, June 2024{\natexlab{a}}. IEEE.
\newblock ISBN 9798350359169.
\newblock \doi{10.1109/ICSA59870.2024.00020}.

\bibitem[Bogner et~al.(2024{\natexlab{b}})Bogner, Wójcik, and Zimmermann]{bogner_how_2024}
Justus Bogner, Pawel Wójcik, and Olaf Zimmermann.
\newblock How {Do} {Microservice} {API} {Patterns} {Impact} {Understandability}? {A} {Controlled} {Experiment}.
\newblock In \emph{2024 {IEEE} 21st {International} {Conference} on {Software} {Architecture} ({ICSA})}, pages 158--169, Hyderabad, India, June 2024{\natexlab{b}}. IEEE.
\newblock ISBN 9798350359169.
\newblock \doi{10.1109/ICSA59870.2024.00023}.

\bibitem[Budgen et~al.(2018)Budgen, Brereton, Drummond, and Williams]{Budgen2018}
David Budgen, Pearl Brereton, Sarah Drummond, and Nikki Williams.
\newblock Reporting systematic reviews: {Some} lessons from a tertiary study.
\newblock \emph{Information and Software Technology}, 95:\penalty0 62--74, March 2018.
\newblock ISSN 09505849.
\newblock \doi{10.1016/j.infsof.2017.10.017}.
\newblock Publisher: Elsevier B.V.

\bibitem[Castaño et~al.(2023)Castaño, Martínez-Fernández, Franch, and Bogner]{castano_exploring_2023}
Joel Castaño, Silverio Martínez-Fernández, Xavier Franch, and Justus Bogner.
\newblock Exploring the {Carbon} {Footprint} of {Hugging} {Face}'s {ML} {Models}: {A} {Repository} {Mining} {Study}.
\newblock In \emph{2023 {ACM}/{IEEE} {International} {Symposium} on {Empirical} {Software} {Engineering} and {Measurement} ({ESEM})}, pages 1--12, New Orleans, LA, USA, October 2023. IEEE.
\newblock ISBN 978-1-66545-223-6.
\newblock \doi{10.1109/ESEM56168.2023.10304801}.

\bibitem[El~Haji et~al.(2024)El~Haji, Brandt, and Zaidman]{el2024using}
Khalid El~Haji, Carolin Brandt, and Andy Zaidman.
\newblock Using github copilot for test generation in python: An empirical study.
\newblock In \emph{Proceedings of the 5th ACM/IEEE International Conference on Automation of Software Test (AST 2024)}, pages 45--55, 2024.

\bibitem[He et~al.(2024)He, Parikh, Weimer, and Endres]{he_high_2024}
Wenxin He, Manasvi Parikh, Westley Weimer, and Madeline Endres.
\newblock High {Expectations}: {An} {Observational} {Study} of {Programming} and {Cannabis} {Intoxication}.
\newblock In \emph{Proceedings of the {IEEE}/{ACM} 46th {International} {Conference} on {Software} {Engineering}}, pages 1--12, Lisbon Portugal, April 2024. ACM.
\newblock ISBN 9798400702174.
\newblock \doi{10.1145/3597503.3639145}.

\bibitem[Hevner(2007)]{hevner_three_2007}
Alan~R Hevner.
\newblock A {Three} {Cycle} {View} of {Design} {Science} {Research}.
\newblock \emph{Scandinavian Journal of Information Systems}, 19\penalty0 (2):\penalty0 87--92, 2007.
\newblock ISSN 09050167.
\newblock ISBN: 0905-0167.

\bibitem[Jedlitschka et~al.(2008)Jedlitschka, Ciolkowski, and Pfahl]{Jedlitschka2008}
Andreas Jedlitschka, Marcus Ciolkowski, and Dietmar Pfahl.
\newblock Reporting {Experiments} in {Software} {Engineering}.
\newblock In \emph{Guide to {Advanced} {Empirical} {Software} {Engineering}}, pages 201--228. Springer London, London, 2008.
\newblock \doi{10.1007/978-1-84800-044-5_8}.

\bibitem[Kitchenham et~al.(2023)Kitchenham, Madeyski, and Budgen]{kitchenham_segress_2023}
Barbara Kitchenham, Lech Madeyski, and David Budgen.
\newblock {SEGRESS}: {Software} {Engineering} {Guidelines} for {REporting} {Secondary} {Studies}.
\newblock \emph{IEEE Transactions on Software Engineering}, 49\penalty0 (3):\penalty0 1273--1298, March 2023.
\newblock ISSN 0098-5589, 1939-3520, 2326-3881.
\newblock \doi{10.1109/TSE.2022.3174092}.

\bibitem[Kozanidis et~al.(2022)Kozanidis, Verdecchia, and Guzm{\'a}n]{kozanidis2022asking}
Nicholas Kozanidis, Roberto Verdecchia, and Emitz{\'a} Guzm{\'a}n.
\newblock Asking about technical debt: Characteristics and automatic identification of technical debt questions on stack overflow.
\newblock In \emph{Proceedings of the 16th ACM/IEEE International Symposium on Empirical Software Engineering and Measurement}, pages 45--56, 2022.

\bibitem[Luccioni et~al.(2024)Luccioni, Jernite, and Strubell]{luccioni_power_2024}
Sasha Luccioni, Yacine Jernite, and Emma Strubell.
\newblock Power {Hungry} {Processing}: {Watts} {Driving} the {Cost} of {AI} {Deployment}?
\newblock In \emph{The 2024 {ACM} {Conference} on {Fairness}, {Accountability}, and {Transparency}}, pages 85--99, Rio de Janeiro Brazil, June 2024. ACM.
\newblock ISBN 9798400704505.
\newblock \doi{10.1145/3630106.3658542}.

\bibitem[Maggi et~al.(2024)Maggi, Verdecchia, Scommegna, and Vicario]{maggi2024claim}
Kevin Maggi, Roberto Verdecchia, Leonardo Scommegna, and Enrico Vicario.
\newblock Claim: a lightweight approach to identify microservices in dockerized environments.
\newblock In \emph{Proceedings of the 28th International Conference on Evaluation and Assessment in Software Engineering}, pages 357--362, 2024.

\bibitem[Maggi et~al.(2025)Maggi, Verdecchia, Scommegna, and Vicario]{maggi2025evolution}
Kevin Maggi, Roberto Verdecchia, Leonardo Scommegna, and Enrico Vicario.
\newblock Evolution of code technical debt in microservices architectures.
\newblock \emph{Journal of Systems and Software}, 222:\penalty0 112301, 2025.

\bibitem[Martínez-Fernández et~al.(2022)Martínez-Fernández, Bogner, Franch, Oriol, Siebert, Trendowicz, Vollmer, and Wagner]{Martinez-Fernandez2022}
Silverio Martínez-Fernández, Justus Bogner, Xavier Franch, Marc Oriol, Julien Siebert, Adam Trendowicz, Anna~Maria Vollmer, and Stefan Wagner.
\newblock Software {Engineering} for {AI}-{Based} {Systems}: {A} {Survey}.
\newblock \emph{ACM Transactions on Software Engineering and Methodology}, 31\penalty0 (2):\penalty0 1--59, April 2022.
\newblock ISSN 1049-331X.
\newblock \doi{10.1145/3487043}.
\newblock arXiv: 2105.01984.

\bibitem[Mendez et~al.(2020)Mendez, Graziotin, Wagner, and Seibold]{Mendez2020}
Daniel Mendez, Daniel Graziotin, Stefan Wagner, and Heidi Seibold.
\newblock Open {{Science}} in {{Software Engineering}}.
\newblock In Michael Felderer and Guilherme~Horta Travassos, editors, \emph{Contemporary {{Empirical Methods}} in {{Software Engineering}}}, pages 477--501. Springer International Publishing, Cham, 2020.
\newblock ISBN 978-3-030-32488-9 978-3-030-32489-6.
\newblock \doi{10.1007/978-3-030-32489-6_17}.

\bibitem[Migliorini et~al.(2024)Migliorini, Verdecchia, Malavolta, Lago, and Vicario]{migliorini2024architectural}
Sofia Migliorini, Roberto Verdecchia, Ivano Malavolta, Patricia Lago, and Enrico Vicario.
\newblock Architectural views: The state of practice in open-source software projects.
\newblock In \emph{European Conference on Software Architecture}, pages 396--415. Springer, 2024.

\bibitem[Miranda et~al.(2018)Miranda, Cruciani, Verdecchia, and Bertolino]{miranda2018fast}
Breno Miranda, Emilio Cruciani, Roberto Verdecchia, and Antonia Bertolino.
\newblock Fast approaches to scalable similarity-based test case prioritization.
\newblock In \emph{Proceedings of the 40th International Conference on Software Engineering}, pages 222--232, 2018.

\bibitem[Rodr{\'\i}guez-P{\'e}rez et~al.(2020)Rodr{\'\i}guez-P{\'e}rez, Robles, Serebrenik, Zaidman, Germ{\'a}n, and Gonzalez-Barahona]{rodriguez2020bugs}
Gema Rodr{\'\i}guez-P{\'e}rez, Gregorio Robles, Alexander Serebrenik, Andy Zaidman, Daniel~M Germ{\'a}n, and Jesus~M Gonzalez-Barahona.
\newblock How bugs are born: A model to identify how bugs are introduced in software components.
\newblock \emph{Empirical Software Engineering}, 25:\penalty0 1294--1340, 2020.

\bibitem[Stol et~al.(2016)Stol, Ralph, and Fitzgerald]{Stol2016}
Klaas-Jan Stol, Paul Ralph, and Brian Fitzgerald.
\newblock Grounded theory in software engineering research.
\newblock In \emph{Proceedings of the 38th {International} {Conference} on {Software} {Engineering} - {ICSE} '16}, pages 120--131, New York, New York, USA, 2016. ACM Press.
\newblock ISBN 978-1-4503-3900-1.
\newblock \doi{10.1145/2884781.2884833}.

\bibitem[Tufano et~al.(2015)Tufano, Palomba, Bavota, Oliveto, Di~Penta, De~Lucia, and Poshyvanyk]{tufano2015and}
Michele Tufano, Fabio Palomba, Gabriele Bavota, Rocco Oliveto, Massimiliano Di~Penta, Andrea De~Lucia, and Denys Poshyvanyk.
\newblock When and why your code starts to smell bad.
\newblock In \emph{2015 IEEE/ACM 37th IEEE International Conference on Software Engineering}, volume~1, pages 403--414. IEEE, 2015.

\bibitem[Verdecchia and Lago(2022)]{verdecchia2022tales}
Roberto Verdecchia and Patricia Lago.
\newblock Tales of hybrid teaching in software engineering: Lessons learned and guidelines.
\newblock \emph{IEEE Transactions on Education}, 66\penalty0 (3):\penalty0 234--243, 2022.

\bibitem[Verdecchia et~al.(2020)Verdecchia, Kruchten, and Lago]{verdecchia2020architectural}
Roberto Verdecchia, Philippe Kruchten, and Patricia Lago.
\newblock Architectural technical debt: A grounded theory.
\newblock In \emph{Software Architecture: 14th European Conference, ECSA 2020, L'Aquila, Italy, September 14--18, 2020, Proceedings 14}, pages 202--219. Springer, 2020.

\bibitem[Verdecchia et~al.(2021)Verdecchia, Kruchten, Lago, and Malavolta]{verdecchia2021building}
Roberto Verdecchia, Philippe Kruchten, Patricia Lago, and Ivano Malavolta.
\newblock Building and evaluating a theory of architectural technical debt in software-intensive systems.
\newblock \emph{Journal of Systems and Software}, 176:\penalty0 110925, 2021.

\bibitem[Verdecchia et~al.(2023)Verdecchia, Engstr{\"o}m, Lago, Runeson, and Song]{verdecchia2023threats}
Roberto Verdecchia, Emelie Engstr{\"o}m, Patricia Lago, Per Runeson, and Qunying Song.
\newblock Threats to validity in software engineering research: A critical reflection.
\newblock \emph{Information and Software Technology}, 164:\penalty0 107329, 2023.

\bibitem[Wieringa(2014)]{Wieringa2014a}
Roel~J. Wieringa.
\newblock Research {Goals} and {Research} {Questions}.
\newblock In \emph{Design {Science} {Methodology} for {Information} {Systems} and {Software} {Engineering}}, pages 13--23. Springer Berlin Heidelberg, Berlin, Heidelberg, 2014.
\newblock ISBN 978-3-662-43839-8.
\newblock \doi{10.1007/978-3-662-43839-8_2}.
\newblock ISSN: 0270-5257.

\bibitem[Wohlin et~al.(2024)Wohlin, Runeson, Höst, Ohlsson, Regnell, and Wesslén]{wohlin_experimentation_2024}
Claes Wohlin, Per Runeson, Martin Höst, Magnus~C. Ohlsson, Björn Regnell, and Anders Wesslén.
\newblock \emph{Experimentation in {Software} {Engineering}}.
\newblock Springer Berlin Heidelberg, Berlin, Heidelberg, 2nd edition, 2024.
\newblock ISBN 978-3-662-69305-6 978-3-662-69306-3.
\newblock \doi{10.1007/978-3-662-69306-3}.

\bibitem[Wyrich and Apel(2024)]{Wyrich2024}
Marvin Wyrich and Sven Apel.
\newblock Evidence {{Tetris}} in the {{Pixelated World}} of {{Validity Threats}}.
\newblock In \emph{Proceedings of the 1st {{IEEE}}/{{ACM International Workshop}} on {{Methodological Issues}} with {{Empirical Studies}} in {{Software Engineering}}}, pages 13--16, Lisbon Portugal, April 2024. ACM.
\newblock ISBN 9798400705670.
\newblock \doi{10.1145/3643664.3648203}.

\bibitem[Yang et~al.(2025)Yang, Jakubowski, Kang, Yu, and Menzies]{yang2025sparsecoder}
Xueqi Yang, Mariusz Jakubowski, Li~Kang, Haojie Yu, and Tim Menzies.
\newblock Sparsecoder: Advancing source code analysis with sparse attention and learned token pruning.
\newblock \emph{Empirical Software Engineering}, 30\penalty0 (1):\penalty0 1--30, 2025.

\bibitem[Zamorano et~al.(2024)Zamorano, Domingo, Cetina, and Sarro]{zamorano2024game}
Mar Zamorano, {\'A}frica Domingo, Carlos Cetina, and Federica Sarro.
\newblock Game software engineering: A controlled experiment comparing automated content generation techniques.
\newblock In \emph{Proceedings of the 18th ACM/IEEE International Symposium on Empirical Software Engineering and Measurement}, pages 302--313, 2024.

\end{thebibliography}

\end{document}